\pdfoutput=1
\documentclass[journal=jpclcd,manuscript=article,layout=twocolumn]{achemso}

\usepackage{notes2bib}
\usepackage[T1]{fontenc} 
\usepackage{mathptmx}
\usepackage{etoolbox}
\usepackage{amsmath}
\usepackage{amssymb}

\usepackage{braket}
\usepackage{color}
\usepackage[hidelinks]{hyperref}
\usepackage{accents}
\usepackage{graphicx}
\usepackage{multirow}

\definecolor{darkgreen}{RGB}{40,110,5}
\usepackage{pgfplotstable}
\usepackage{rotating}
\usepackage{tabularx,booktabs}
\usepackage{adjustbox}
\usepackage{sidecap}
\usepackage{epstopdf}
\usepackage{gensymb}
\usepackage{float}

\DeclareUnicodeCharacter{2212}{-}
\usepackage{color}
\usepackage{ulem}
\definecolor{darkgreen}{RGB}{40,110,5}

\newcommand{\onlinecite}[1]{\hspace{-1 ex} \nocite{#1}\citenum{#1}}

\title[]{Characterizing the Role of Peierls Vibrations in Singlet Fission with the Adaptive Hierarchy of Pure States}

\author{Jacob K. Lynd}%
\affiliation{ 
Department of Chemistry, University of Texas at Austin, Austin, TX, 78712, USA
}%
 
\author{Doran I.G.B. Raccah}
\affiliation{ 
Department of Chemistry, University of Texas at Austin, Austin, TX, 78712, USA
}%
 \email{doran.raccah@utexas.edu}
\date{\today}

\abbreviations{IR,NMR,UV}

\begin{document}
\sloppy 


\begin{tocentry}

\includegraphics[width=\linewidth]{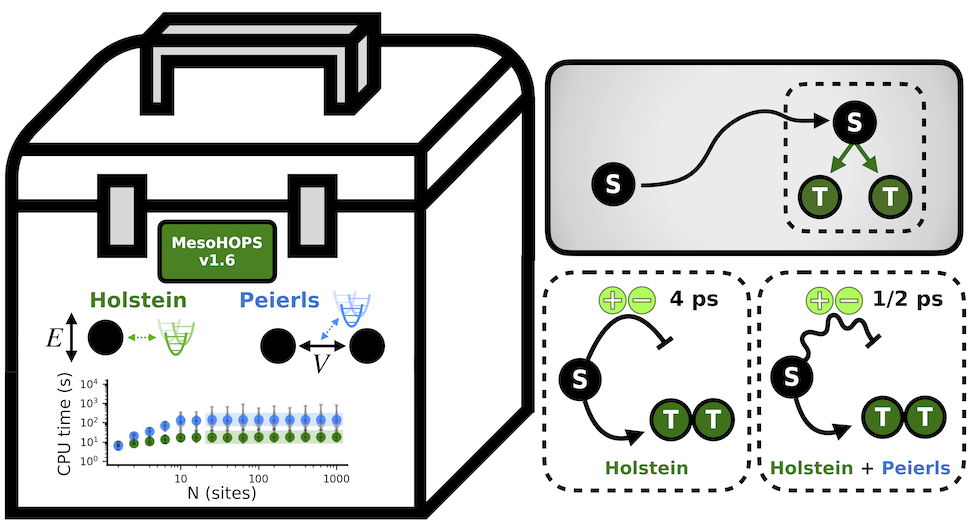}





\end{tocentry}


\begin{abstract}
\label{LH2:abs}
  Singlet fission, a phenomenon in which a singlet exciton is converted to two triplet excitons, is sensitive to vibrations that perturb couplings between electronic states (i.e., Peierls vibrations). In singlet fission models larger than dimers, the inconvenient scaling of exact simulations has limited treatment of Peierls vibrations to approximate methods. In this letter, we generalize the formally exact, reduced-scaling adaptive Hierarchy of Pure States (adHOPS) method to account for both Holstein and Peierls vibrations and study singlet fission in N,N'-Bis(2-phenylethyl)-3,4,9,10-perylenedi-carboximide (EP-PDI). We find that Peierls vibrations accelerate singlet fission by generating correlated charge transfer-mediated pathways that support constructive interference. Finally, we extend this singlet fission model to a linear chain of EP-PDI to demonstrate that Peierls vibrations can accelerate singlet-mediated triplet transport on the 100-nm scale.
\end{abstract}

\twocolumn

Excited-state dynamics in molecular materials depend on strong electron-phonon couplings to both intramolecular and intermolecular vibrational degrees of freedom.\cite{Fratinipeierls2009, Nematiaram2020Parameterizations,Li_OSC_2021} Holstein vibrations, usually associated with intramolecular degrees-of-freedom, drive fluctuations in excited-state energies,\cite{Holstein1959} whereas primarily-intermolecular Peierls vibrations augment electronic couplings between states.\cite{Girlandorev2010,SuSchriefferHeeger1979} The interplay between the two classes of electron-phonon interactions presents a rich space of physics, but the enormous manifold of parameters to be explored demands accurate and flexible methodologies for meaningful simulations.\cite{Nematiaram2020Parameterizations, BerkelbachUnified2020,Li_OSC_2021,Zhou_2024_HOMPS_Peierls}

In particular, understanding the role of Peierls vibrations is vital for constructing accurate models of singlet fission (SF). SF is a spin-allowed process in which a singlet exciton on an organic semiconductor molecule splits into two neighboring triplet excitations.\cite{Smith_2010_SF_Review} Recent organic photovoltaic technologies have taken advantage of the ensuing greater-than-unity quantum yield of photoexcited charge carriers to increase output current\cite{takeda2024, nagaya2025} and can theoretically go beyond the Shockley-Queisser limit to achieve power conversion efficiencies nearing 50\%.\cite{BerkelbachSFI} Previous simulations and experiments have indicated that intermolecular vibrations play a vital role in SF dynamics.\cite{Renaud2015SFPeierls,Duan2020SFPeierls,Castellanos2017SFPeierls,Tempelaar2018SFPeierls,Miyata_2017_Symmetry_Breaking, alvertis2019solventpolarity,tamura2015pistacking,Alguire2015noncondon} In this letter, we simulate SF using the reduced-scaling and formally exact adaptive Hierarchy of Pure States (adHOPS) to show that Peierls vibrations induce symmetry-breaking fluctuations in electronic couplings, accelerating SF via constructive interference.

We describe molecular materials as open quantum systems, where the electronic degrees of freedom are represented explicitly and the vibrational degrees of freedom are treated as a set of thermal reservoirs. The energies and couplings of the electronic states $\ket{i}$ comprise the system Hamiltonian
\begin{equation}
    \hat{H}_S = \sum_i E_i|i\rangle\langle i| + \sum_{j}\sum_{j<i} V_{ij}|i\rangle\langle j| + h.c.
\end{equation}
and the vibrational environment is described by a set of independent harmonic baths, indexed $n$, where each mode $q_n$ (defined by frequency $\omega_{q_n}/\hslash$ and creation operator $\hat{a}_{q_n}^\dag$) is linearly coupled to the system with strength $\Lambda_{q_n}$ by system-bath coupling operator $\hat{L}_n$. The full Hamiltonian is thus
\begin{equation}
\label{eq:open_quantum_system_hamiltonian}
    \hat{H} = \hat{H}_S + \sum_{n,q_n} \hat{L}_n\Lambda_{q_n}\big(\hat{a}^\dag_{q_n} + \hat{a}_{q_n}\big) + \sum_{n,q_n} \omega_{q_n}\big(\hat{a}^\dag_{q_n}\hat{a}_{q_n} + 1/2\big).
\end{equation}
Each bath's couplings to the system are described by a spectral density,
\begin{equation}
    J_n(\omega) = \pi\sum_{q_n}|\Lambda_{q_n}|^2\delta(\omega-\omega_{q_n}).
\end{equation}
The time correlation function of fluctuations in $\hat{H}_S$ induced by the bath at inverse temperature $\beta=\frac{1}{k_BT}$,\cite{may_kuhn,Feynman_student_thesis}
\begin{equation}
\label{eq:C_t}
C_n(t) = \frac{1}{\pi}\int_0^\infty d\omega J_n(\omega) \big(\coth(\frac{\beta\omega }{2}) \cos(\omega t/\hslash) - i \sin( \omega t/\hslash)\big)
\end{equation}
may be decomposed into a sum of complex exponential modes
\begin{equation}
\label{eq:exp_modes}
    C_n(t) = \sum_{j_n}C_{j_n}(t) = \sum_{j_n}g_{j_n}e^{-\gamma_{j_n}t/\hslash}
\end{equation}
where $\textrm{Re}[\hslash/\gamma_{j_n}]$ is the timescale of vibrational relaxation.

The Hierarchy of Pure States (HOPS) method propagates the exact excited-state dynamics of open quantum systems.\cite{HOPS} HOPS uses a Monte-Carlo sampling over coherent states of the harmonic bath to produce a set of independent trajectories indexed by a stochastic process $\mathbf{z}$, comprised of terms $z_{n,t}$ for the $n^{\textrm{th}}$ bath at time $t$ that satisfy the ensemble average $\mathbb{E}_\mathbf{z}[z^*_{n,t}z_{m,s}] = \delta_{m,n}C_n(t-s)$.\cite{HOPS, NMQSD} The system density matrix at time $t$,\cite{NMQSD,HOPS}
\begin{equation}
    \hat{\rho}_t = \mathbb{E}_\mathbf{z}[|\psi_t^{(\vec{0})}(\mathbf{z})\rangle\langle\psi_t^{(\vec{0})}(\mathbf{z})|]
\end{equation} 
is reproduced by the ensemble average of system wave functions $|\psi_t^{(\vec{0})}(\mathbf{z})\rangle$ across all trajectories.
HOPS accounts for the non-Markovian bath memory with an equation-of-motion (given in section S1.B of the SI) that couples  $|\psi_t^{(\vec{0})}(\mathbf{z})\rangle$ to a high-dimensional grid (hierarchy) of auxiliary wave functions $|\psi_t^{(\vec{k})}(\mathbf{z})\rangle$, indexed by vector $\vec{k}$ in the space of exponential correlation function modes (Eq. \eqref{eq:exp_modes}) and truncated by the triangular condition $\|\vec{k}\|_1 \leq k_{\textrm{max}}$. Thus, despite convenient statistical convergence\cite{HOPS} and several powerful extensions,\cite{2noiseHOPS, nuHOPS, gaoNonMarkovianStochasticSchrodinger2022, Zhou_2024_HOMPS_Peierls, boettcher_2024_bath_observables, chen_2022_NLR} HOPS calculations scale catastrophically with respect to system size because the dimensionality of the hierarchy scales with the number of independent vibrational baths.

\begin{figure}[h!]
    \centering
    \includegraphics[width=\linewidth]{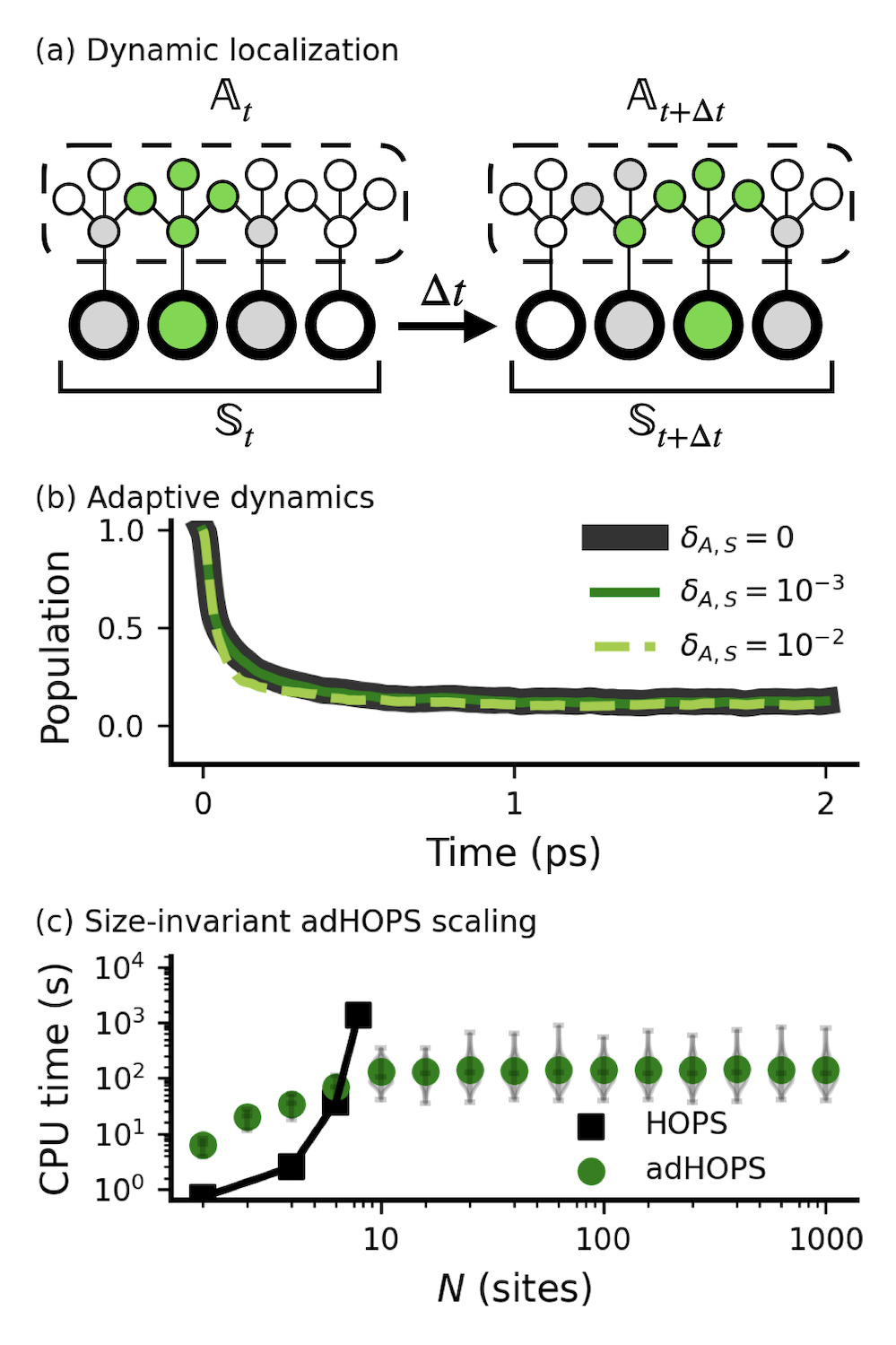}
    \caption{The generalized adHOPS exhibits size-invariance in formally exact simulations of a Peierls model. (a) The dynamic localization of excitation (green) in the HOPS framework ensures that the number of relevant (green and grey) electronic states (large circles) and auxiliary wave functions (small circles) at each point in time depends on delocalization extent rather than size of the full system. (b) Excited-state dynamics following excitation of the leftmost site in a 7-site linear chain calculated by HOPS (black) and adHOPS with various adaptive error bounds (green). (c) Distribution over trajectories of the CPU time to propagate 1 ps of exciton dynamics at various chain lengths $N$ using adHOPS with $\delta_A=\delta_S=10^{-3}$ (green) compared to mean HOPS CPU time (black). See Table S3 of the SI for calculation details.}
    \label{fig:peierls_adaptivity}
\end{figure}

The adaptive Hierarchy of Pure States method\cite{adHOPS} (adHOPS) takes advantage of localization to reduce the difficulty of HOPS calculations in extended systems.\cite{ varvelo2023formally,gera_simulating_2023,MesoHOPS2024,gera2025fluorescence} In the HOPS formalism, strong system-bath interactions localize excitations, ensuring that the majority of electronic states and auxiliary wave functions are unpopulated (and thus irrelevant)  at any given time $t$ in each trajectory, so long as both electronic coupling length scales and bath reorganization times are finite.\cite{adHOPS} The adHOPS algorithm quantifies the error that excluding a given auxiliary wave function or system state introduces in the derivative, finding an upper bound under the assumption that system-bath coupling operators are diagonal ($\hat{L}_n = \sum_i l_i|i\rangle\langle i|$). \cite{MesoHOPS2024} The algorithm removes auxiliary wave functions from the calculation from smallest to largest error while obeying a user-defined error bound, $\delta_A$, before doing the same for system states with error bound $\delta_S$. The ensuing reduced calculations remain formally exact, with $\delta_A$ and $\delta_S$ acting as convergence parameters.\cite{adHOPS, varvelo2023formally, gera_simulating_2023, MesoHOPS2024, gera2025fluorescence} As the excitation propagates through the material, once-populated electronic states and auxiliary wave functions become unpopulated, as shown schematically in Fig. \ref{fig:peierls_adaptivity}a, and the sizes of the adaptive state ($\mathbb{S}_t$) and auxiliary ($\mathbb{A}_t$) bases depend only on delocalization length. Because delocalization of excitations does not depend on the full extent of a material with finite coupling length, adHOPS achieves size-invariant ($\mathcal{O}(1)$) scaling in large systems.\cite{adHOPS,MesoHOPS2024}

In this letter, we introduce a new version of adHOPS that accounts for both Holstein and Peierls-type electronic-vibrational couplings. The previous requirement that system-bath coupling operators take the diagonal form $\hat{L}_n = \sum_i l_i|i\rangle\langle i|$ limited adHOPS to systems with only Holstein vibrations. In section S2 of the SI, we detail a new adHOPS algorithm that accounts for the generalized Hermitian form $\hat{L}_n = \sum_{i}L_n(i,i)|i\rangle \langle i| + \sum_{j< i} (L_n(i,j)|i\rangle \langle j| + h.c.)$ that encompasses both Holstein and Peierls vibrations. This updated adHOPS algorithm is available in MesoHOPS version 1.6. MesoHOPS depends on the NumPy, SciPy, and Numba libraries, and has an installation guide and introductory tutorial on \href{https://github.com/MesoscienceLab/mesohops}{GitHub}.

The new version of adHOPS remains exact and exhibits size-invariance in systems with Peierls vibrations. We simulate exciton dynamics in a linear chain comprised of $N$ identical pigments with nearest-neighbor electronic couplings such that
\begin{equation}
\label{eq:linearchainhamiltonian}
    \hat{H}_S = \sum_{n=1}^{N} V\ket{n}\bra{n+1} +h.c.
\end{equation}
The coupling between each pair of nearest-neighbor sites is modulated by a Peierls bath with system-bath coupling operator
\begin{equation}
    \hat{L}_n = \ket{n}\bra{n+1} + h.c.
\end{equation}
Each bath is described by a Drude-Lorentz spectral density ($J_n(\omega) = 2\lambda_n\gamma_n\frac{\omega}{\omega^2 + \gamma_n^2}$) with reorganization energy $\lambda_n$ and timescale $\hslash/\gamma_n$. The model is parametrized by $\lambda_n=\gamma_n=V=50\textrm{ cm}^{-1}$ at a temperature of $300$ K.
In Fig. \ref{fig:peierls_adaptivity}b, we show that as $\delta_A$ and $\delta_S$ decrease, the population dynamics of the initially-excited state in a linear chain with $N=7$ converge to the non-adaptive $\delta_A=\delta_S=0$ case. In Fig. \ref{fig:peierls_adaptivity}c, we show that adHOPS exhibits reduced scaling, reaching size-invariance by $N=30$. Comparatively, a non-adaptive HOPS calculation becomes prohibitively expensive around $N=10$.\bibnote{All CPU times were measured by running each trajectory on a single thread of an AMD EPYC 9534 64-Core Processor.} The new version of adHOPS retains the accuracy and convenient scaling of previous versions while being applicable to systems with both Holstein and Peierls vibrations.

\begin{figure}[h!]
    \centering
    \includegraphics[width=\linewidth]{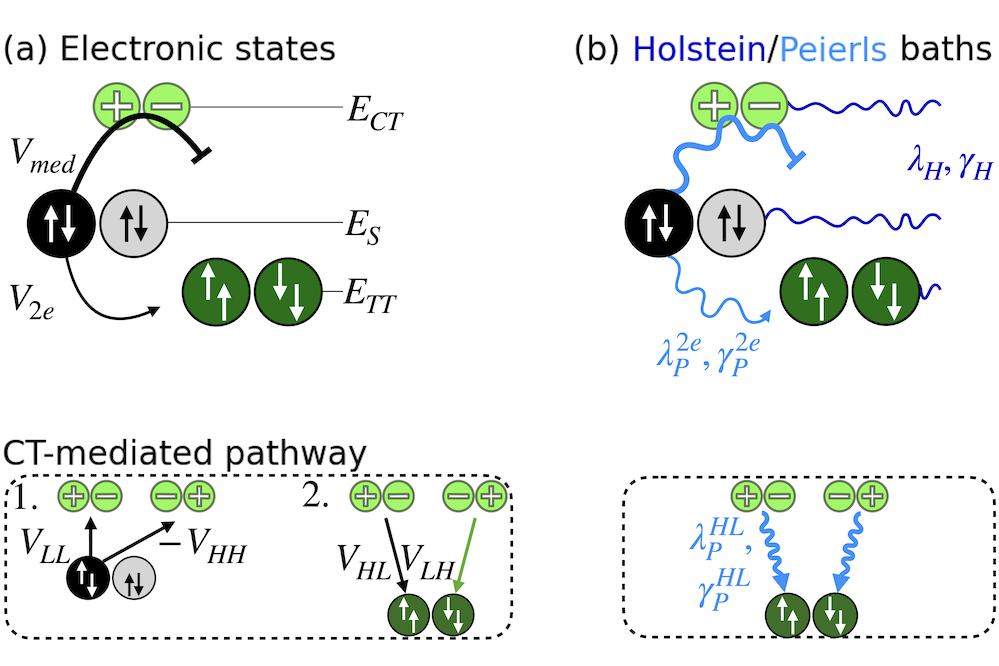}
    \caption{Peierls vibrations may augment both the direct and CT-mediated SF mechanisms in EP-PDI. (a) We include singlet states $\ket{e_n}$ with energy $E_S$ (black), CT states $\ket{A_nC_{n+1}}$/$\ket{C_nA_{n+1}}$ with energy $E_{CT}$ (bright green), and TT states $\ket{T_nT_{n+1}}$ with energy $E_{TT}$ (dark green) in our model. The singlet and TT states are coupled directly ($V_{2e}$) and by a CT-mediated effective coupling ($V_{med}$); the dashed box shows the individual couplings that contribute to $\hat{V}_{\textrm{med}}$, where green arrows indicate negative couplings. (b) Holstein (dark blue) vibrational baths with reorganization timescales $\hslash/\gamma_H$ are coupled with reorganization energy $\lambda_H$ to each state, causing fluctuations in the vertical excitation energies (shown as wiggly lines). Peierls (light blue) vibrational baths (similarly parametrized $\gamma_P,\lambda_P$) may alter the direct couplings between the singlet and TT states or the couplings between CT and TT states (and thus the CT-mediated effective couplings), as shown in the dashed box.}
    \label{fig:hamiltonian}
\end{figure}

We probe the mechanistic role of Peierls vibrations in singlet fission using a perylene di-imide (PDI) linear chain model. PDI derivatives are photostable, readily absorb light, and exhibit fast SF, making them promising candidates for photovoltaics and excellent model systems for exploring SF.\cite{Le_2018_SF_Interplay} Following Refs. \onlinecite{BerkelbachSFI} and \onlinecite{BerkelbachSFIII}, we define the diabatic electronic states based on the highest occupied molecular orbital (HOMO) and lowest unoccupied molecular orbital (LUMO) of each molecule. Singlet states localized on the $n^{\textrm{th}}$ chromophore, $\ket{e_n}$, are coupled to high-energy charge-transfer (CT) states $\ket{C_nA_{n+1}}$ and $\ket{A_nC_{n+1}}$ (and/or $\ket{C_{n-1}A_n}$ and $\ket{A_{n-1}C_n}$) by 1-electron LUMO-LUMO and HOMO-HOMO couplings $V_{LL}$ and $V_{HH}$, as shown in Fig. \ref{fig:hamiltonian}a.\bibnote{For the CT state indexing, $C_n$ indicates a cationic hole in the HOMO of molecule $n$ and $A_m$ indicates an anionic electron in the LUMO of molecule $m$.} The CT states localized on sites $n,n+1$ are, in turn, coupled to the triplet pair (TT) state $\ket{T_nT_{n+1}}$ by HOMO-LUMO couplings $V_{HL}=-V_{LH}$, which are anti-symmetric due to orbital geometries.\cite{Castellanos2017SFPeierls} States of all three types participate in a CT-mediated mechanism of SF wherein the high-energy CT states act as a transiently-populated intermediate between the singlet and TT states, characterized by an effective coupling $V_{med}$ (given in section S6.B of the SI).\cite{Castellanos2017SFPeierls} There is also a weak direct 2-electron coupling between $\ket{e_n}$ and $\ket{T_nT_{n+1}}$ (and/or $\ket{T_{n-1}T_n}$). Our model includes both Holstein  and Peierls vibrations, the effects of which are illustrated in Fig. \ref{fig:hamiltonian}b: each electronic state has an independent Holstein bath (dark blue) that introduces time-dependence to the excited-state energies, and Peierls vibrations (light blue) may augment either the singlet-TT direct couplings or the CT-TT HOMO-LUMO couplings. Several key parameters of our model, based on an N,N'-Bis(2-phenylethyl)-3,4,9,10-perylenedi-carboximide (EP-PDI) dimer at room temperature (Ref. \onlinecite{Renaud2015SFPeierls}), are given in Table \ref{tab:PDI_key_params}. In-depth parameters are given in Table S2 of the SI, along with a detailed Hamiltonian in section S5.

\begin{table}[h!]
    \centering
    \begin{tabular}{|c|c|}
    \hline
        Parameter & Value (cm$^{-1}$)  \\ \hline
         $E_S-E{TT}$ & 2200 \\         \hline

         $E_{CT}-E_{TT}$ & 9800 \\         \hline

         $V_{2e}$ & -3.39 \\         \hline

         $|V_{HH}|,|V_{LL}|,|V_{HL}|,|V_{LH}|$ & [968,1170] \\         \hline

         $\lambda_{H}$
         & 621 \\         \hline

         $\lambda_{P}^{2e}$ & 0.726 \\         \hline

         $\lambda_{P}^{HL}$ & 217 \\         \hline
    \end{tabular}
    \caption{Key EP-PDI parameters.}
    \label{tab:PDI_key_params}
\end{table}

\begin{figure}[h!]
    \centering
    \includegraphics[width=\linewidth]{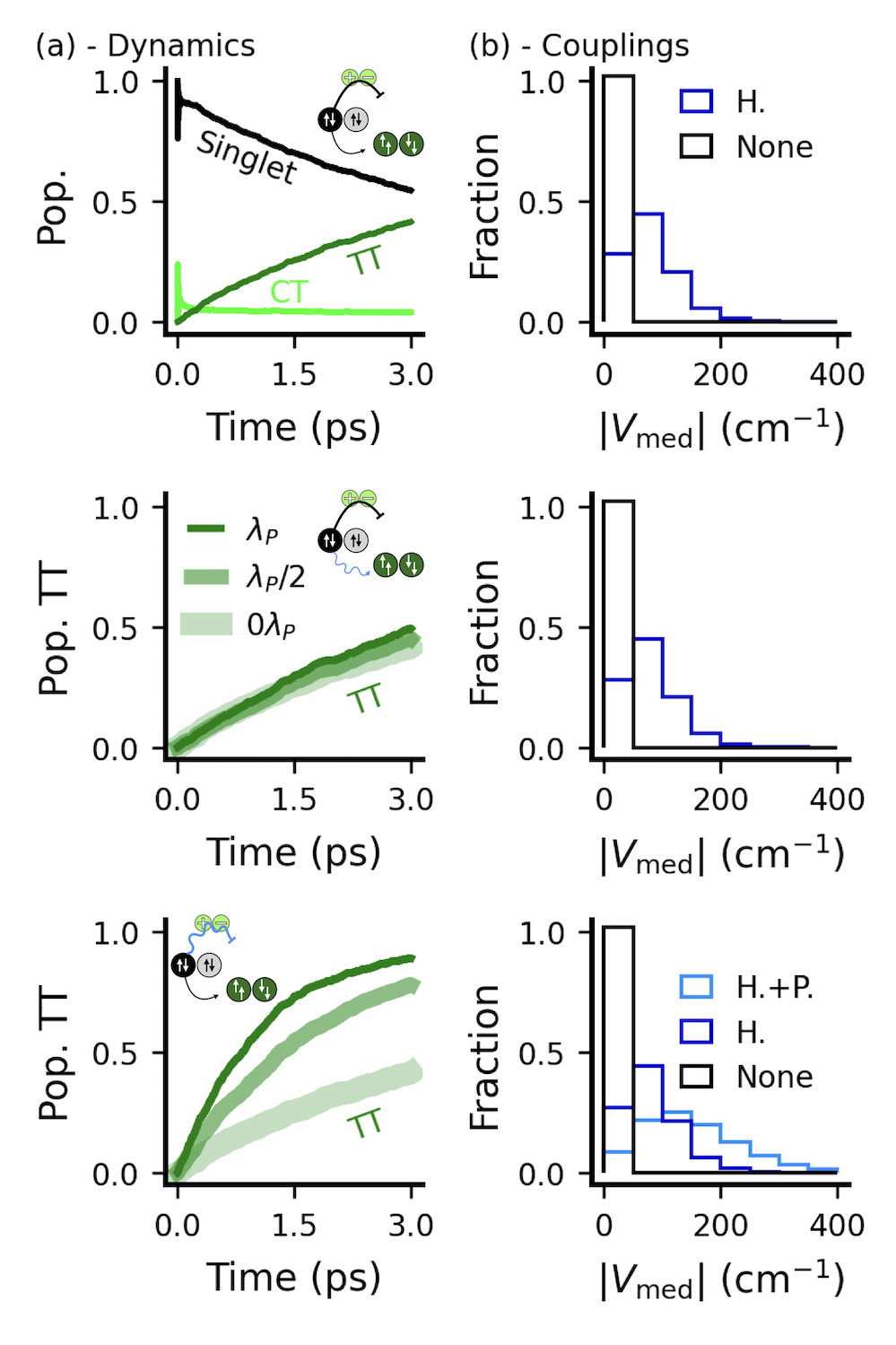}
    \caption{SF in an EP-PDI dimer is accelerated when Peierls vibrations augment the CT-mediated mechanism. The top, middle, and bottom rows represent dimers with no Peierls vibrations, Peierls vibrations modulating the direct singlet-TT electronic coupling, and Peierls vibrations modulating the CT-TT couplings, respectively. (a) Population dynamics of singlet (black), CT (lime), and TT (dark green) states in the absence of Peierls vibrations (top). Population dynamics of TT states with Peierls vibrations at increasing system-bath coupling strengths (darker, thinner lines) (middle and bottom). (b) Time-dependent magnitudes of effective couplings $\braket{e_1|\hat{V}_{\textrm{med}}|T_1T_2}$ and $\braket{e_2|\hat{V}_{\textrm{med}}|T_1T_2}$ sampled across 1000 trajectories every 10 fs for 500 fs calculated without noise (black), with only the portion of the noise corresponding to Holstein vibrations only (dark blue), and with the full noise (light blue).  See Table S3 of the SI for calculation details.}
    \label{fig:pdi_dimer}
\end{figure}

Simulations of an EP-PDI dimer indicate that the CT-mediated mechanism is the primary driver of SF. In Fig. \ref{fig:pdi_dimer}a, we explore the dynamics with no Peierls vibrations (top row), and find that the rate of triplet (dark green) formation is $\sim 0.25$ ps$^{-1}$, with CT states (bright green) only transiently populated as a result of their high energies. In the middle row, we show that augmenting the direct coupling between S and TT states by adding Peierls vibrations with increasing system-bath coupling strengths (middle row, darker and thinner lines) does little to accelerate SF. Consistent with SF proceeding through the CT-mediated mechanism, even a $\sim6$-fold increase in average magnitude of direct coupling increases the rate of SF by only $\sim$5\% (darkest line). Conversely, Peierls vibrations that affect the couplings characteristic of the CT-mediated mechanism significantly increase the rate of triplet formation (bottom row), showing that the CT-mediated mechanism plays a central role in SF.

Electronic-vibrational couplings accelerate the CT-mediated pathway of SF by breaking anti-symmetry in CT-TT couplings, reducing destructive interference. Following an analysis by Castellanos and Huo,\cite{Castellanos2017SFPeierls} we use the augmented system Hamiltonian that accounts for the time-dependent influence of the vibrational environment (see section S1.C of the SI) in the first-order perturbation theory calculation of $\hat{V}_{\textrm{med}}$ and show the distribution of the effective couplings $\braket{e_n|\hat{V}_{\textrm{med}}|T_nT_{n+1}}$ and $\braket{e_{n+1}|\hat{V}_{\textrm{med}}|T_nT_{n+1}}$ in Fig. \ref{fig:pdi_dimer}b. In the absence of vibrations, the anti-symmetry $V_{HL}=-V_{LH}$ limits the  mediated couplings $V_{\textrm{med}}$ to $\sim 50$ cm$^{-1}$ (black). Holstein vibrations introduce transient asymmetries in the vertical excitation energies of states $\ket{A_nC_{n+1}}$ and $\ket{C_nA_{n+1}}$, increasing the average magnitude of $V_{\textrm{med}}$ (dark blue). The Peierls vibrations that augment the direct coupling $\braket{e_n|\hat{H}_S|T_nT_{n+1}}$ do not influence the effective mediated coupling (middle row). In the bottom row, we show that Peierls vibrations that introduce fluctuations in CT-TT couplings, directly breaking the anti-symmetry $V_{HL}=-V_{LH}$, have a dramatic effect on the CT-mediated mechanism, tripling the mean absolute coupling (bright blue) and increasing SF rates.

\begin{figure}[h!]
    \centering
    \includegraphics[width=\linewidth]{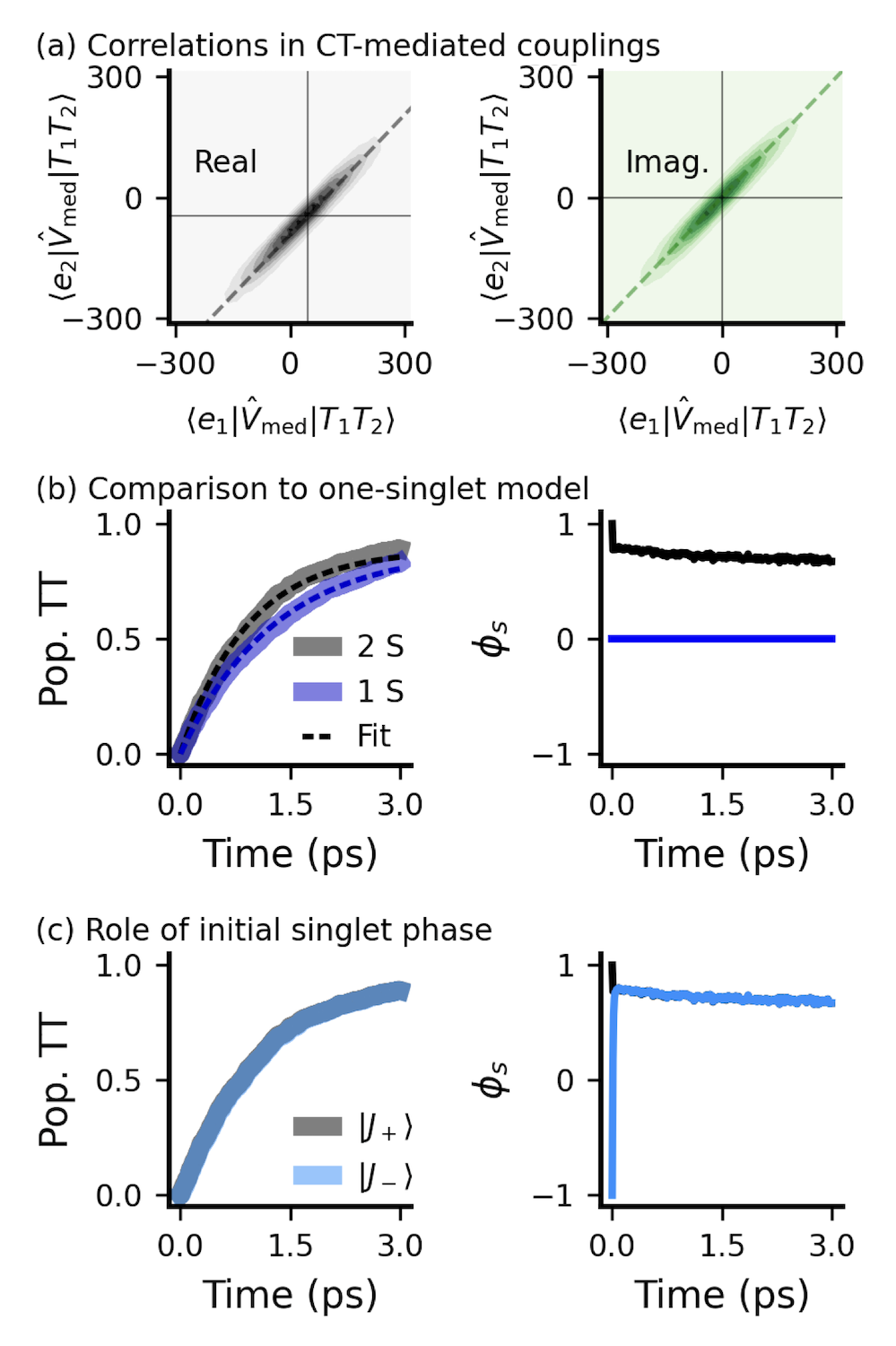}
    \caption{Correlated effective CT-mediated couplings accelerate SF by constructive interference. (a) Correlations in the real (black) and imaginary (green) part of $\braket{e_1|\hat{V}_{\textrm{med}}|T_1T_2}$ and $\braket{e_2|\hat{V}_{\textrm{med}}|T_1T_2}$ sampled across 1000 trajectories every 10 fs for 500 fs in the dimer model presented in the bottom row of Fig. \ref{fig:pdi_dimer}. Thin black lines represent $\hat{V}_{\textrm{med}}$ in the absence of noise. The dashed lines correspond to perfectly correlated couplings. (b) Comparison of the dimer (black) to a truncated model with only one singlet state (dark blue) in terms of triplet population (left) and the ensemble average of $\phi_s$ (right). Dashed lines represent an exponential fit, $P_{TT}(t) = P_{TT}^{eq}\big(1-e^{-kt}\big)$. (c) Comparison of the dimer with initial states $\ket{\psi_0^{(\vec{0})}}=\ket{J_+}$ (black) and $\ket{J_-}$ (light blue) in terms of triplet population (left) and the ensemble average of $\phi_s$. See Table S3 of the SI for calculation details.}
    \label{fig:dimer_phase}
\end{figure}

Uncorrelated vibrational baths induce correlated CT-mediated couplings of a neighboring pair of singlets to their associated TT state. In Fig. \ref{fig:dimer_phase}a, we plot the distributions of the real and imaginary parts of $\braket{e_n|\hat{V}_{\textrm{med}}|T_nT_{n+1}}$ and $\braket{e_{n+1}|\hat{V}_{\textrm{med}}|T_nT_{n+1}}$ in the presence of Holstein and Peierls vibrations (as in the bottom row of Fig. \ref{fig:pdi_dimer}). Without vibrations, these two couplings have the same magnitude and opposite sign (shown as thin black lines). With vibrations, the distributions of both the real and imaginary parts of the coupling are correlated, indicating that the magnitudes and the phases of $\braket{e_n|\hat{V}_{\textrm{med}}|T_nT_{n+1}}$ and $\braket{e_{n+1}|\hat{V}_{\textrm{med}}|T_nT_{n+1}}$ are correlated. This correlation arises because the effective coupling of both singlet states, $\ket{e_n}$ and $\ket{e_{n+1}}$, to $\ket{T_nT_{n+1}}$ is limited by the anti-symmetric coupling of the same pair of CT states to $\ket{T_nT_{n+1}}$. When the anti-symmetry $V_{HL}=-V_{LH}$ is broken by vibrations, mediated couplings of $\ket{e_n}$ and $\ket{e_{n+1}}$ to $\ket{T_nT_{n+1}}$ increase simultaneously.

Correlated effective CT-mediated couplings support coherent effects that accelerate SF. When $\braket{e_{n+1}|\hat{V}_{\textrm{med}}|T_nT_{n+1}}\approx\braket{e_n|\hat{V}_{\textrm{med}}|T_nT_{n+1}}$, a singlet state delocalized in-phase across $\ket{e_n}$ and $\ket{e_{n+1}}$ is strongly coupled to $\ket{T_nT_{n+1}}$ due to constructive interference: the effective coupling is a factor of $\sqrt{2}$ greater than it would be for localized singlet state 
$\ket{e_n}$.
In Fig. \ref{fig:dimer_phase}b, we show that, in a dimer model where Peierls vibrations modify CT-TT couplings, removing the singlet state $\ket{e_2}$ so that the singlet excitation is always localized to a single pigment reduces the rate of triplet formation by $\sim30$\% (dark blue line).  The faster triplet formation when both singlets are available indicates that constructive interference accelerates SF.

Rapid vibrationally-mediated excitonic relaxation ensures that constructive interference accelerates SF, regardless of the initial electronic state. Previous simulations of pentacene indicate that phase effects in the ground state of the singlet manifold control SF rates.\cite{zang_interference_2017} In the right panel of Fig \ref{fig:dimer_phase}b, we plot the ensemble average of 
\begin{equation}
    \phi_s = 2\left|\frac{\braket{\psi^{(\vec{0})}_t|J_+}}{\braket{\psi^{(\vec{0})}_t|(\ket{e_1}\bra{e_1}+\ket{e_2}\bra{e_2})|\psi^{(\vec{0})}_t}}\right|-1
\end{equation}
which has a value of 0 when the singlet excitation is localized and $\pm1$ when the singlet excitation is in delocalized state $\ket{J_\pm}=\frac{1}{\sqrt{2}}\Big(\ket{e_1}\pm\ket{e_2}\Big)$. The positive value of $\phi_s$ when both singlets are available is consistent with constructive interference accelerating SF. In Fig. \ref{fig:dimer_phase}c, we show that $\phi_s$ relaxes to a steady positive value at early times, even when the initial state of the dimer is $\ket{J_-}$ (light blue line). This is consistent with rapid excitonic relaxation to the singlet ground state in a J-aggregate with an effective nearest-neighbor singlet coupling $V_{NN}<0$ (see section S6.A of the SI).\bibnote{We note here that, given these vibrational relaxation effects, an equivalent material in an H-aggregate structure with a positive effective nearest-neighbor coupling $V_{NN}>0$ would have $\ket{J_1}$ as the singlet exciton ground state and the anti-symmetry $V_{HL}=-V_{LH}$ would promote, rather than impede, SF (results not shown).}

\begin{figure*}[h!]
    \centering
    \includegraphics[width=\linewidth]{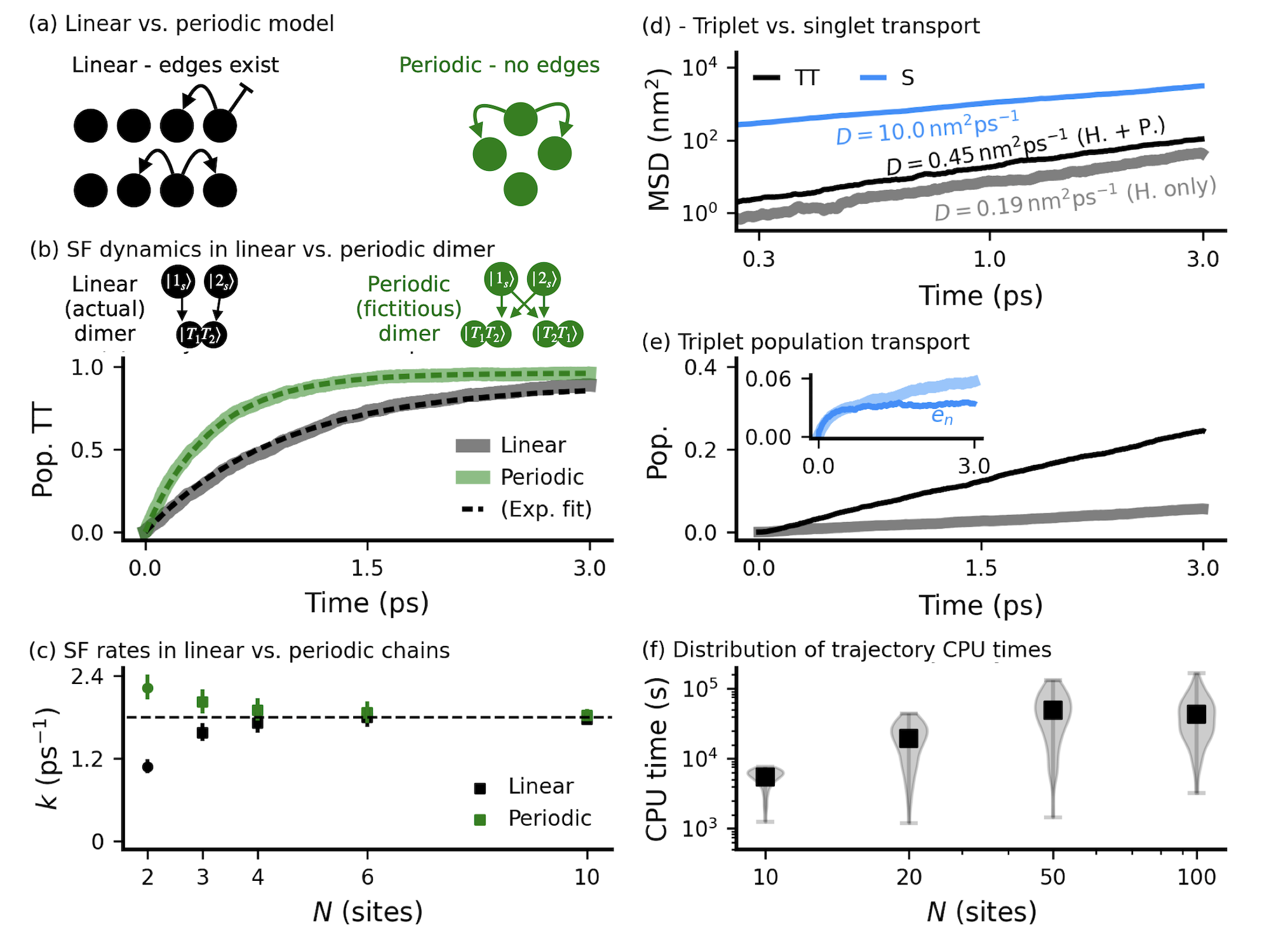}
    \caption{SF and triplet transport dynamics in longer EP-PDI chains. (a) Edge and non-edge molecules in the linear chain model (black) are coupled to different numbers of TT states; a periodic model (green) contains no edge molecules. (b) SF dynamics of a linear and (fictitious) periodic dimer. Dashed lines represent an exponential fit, $P_{TT}(t) = P_{TT}^{eq}\big(1-e^{-kt}\big)$. (c) Rate $k$ of SF in the linear chain and periodic model for different values of $N$ to 99\% confidence. Circles represent dimers. (d) Mean-squared diffusion (see section S5.A of the SI for details) of singlet excitons (blue) and triplet (black) pairs with (thinner, darker) and without (thicker, fainter) Peierls vibrations. We parametrize the transport beyond 1 ps as $\textrm{MSD}(t) = Dt+c$, as determined by a linear best-fit. (e) Populations of triplet states other than $\ket{T_1T_2}$ (black) and singlet states (blue, inset) with (thinner, darker) and without (thicker, fainter) Peierls vibrations. (f) CPU time scaling of the linear chain model at various chain lengths. See Table S3 of the SI for calculation details.}
    \label{fig:PDI_chain}
\end{figure*}

SF rates in molecular materials are sensitive to edge effects and cannot be extracted from a dimer model. In linear chains with a small $N$, the predominance of edge molecules limits SF rates. Singlet state $\ket{e_n}$ can undergo SF to two TT states by exciting either the molecule immediately to the right or to the left, but the leftmost and rightmost singlet states $\ket{e_1}$ and $\ket{N_s}$ are only coupled to a single TT state, as shown in Fig. \ref{fig:PDI_chain}a (black). Conversely, a periodic model with additional triplet state $\ket{T_{N}T_1}$ and corresponding CT states has no edge molecules (\ref{fig:PDI_chain}a, green), but when $N$ is small the increased density of effective singlet-singlet couplings amplifies coherent effects. Thus, as shown in Fig. \ref{fig:PDI_chain}b, the SF rate $k$ is twice as fast in a (fictitious - see section S5 of the SI) periodic dimer. We show in Fig. \ref{fig:PDI_chain}c that as $N$ increases, the two models become indistinguishable, with the $k$ of the linear chain model (black squares) increasing due to a decreasing fraction of edge molecules and the $k$ of the periodic model (green circles) decreasing until they reach agreement at a shared asymptote of $k\approx1.7$ ps$^{-1}$ (dashed line)  at $N\approx 4$.

How might Peierls vibrations influence triplet transport in an EP-PDI linear chain? As triplet mobility is a key factor in enhancing power conversion efficiency of singlet-fission materials,\cite{Volek_2024_SF_TF_Rates} better understanding the role of Peierls vibrations is vital for design of future organic photovoltaics. Recent spectroscopic observations have suggested that in at least some singlet fission materials, like EP-PDI, triplet mobility is accelerated by a singlet mediated mechanism: the triplet pair recombines into a singlet state, the singlet diffuses and then subsequently splits into a new triplet pair.\cite{Volek_2024_SF_TF_Rates,zhu2018exciton} Previous models of this mechanism, however, are parametrized by rate equations that are agnostic to the role of molecular vibrations.

In our EP-PDI model, the presence of Peierls vibrations accelerates singlet-mediated triplet diffusion. The model Hamiltonian presented here limits triplet mobility to the singlet-mediated mechanism due to an absence of direct coupling between triplet pair states. Initializing the excitation in the leftmost triplet pair state $|T_1T_2\rangle$ results in a triplet transport rate (black line, Fig.~\ref{fig:PDI_chain}d) about 1/20$^{\textrm{th}}$ the rate of singlet diffusion (blue, Fig.~\ref{fig:PDI_chain}d, see section S5 of the SI) in a 100-site linear chain. In the absence of the Peierls vibrations, the rate of triplet diffusion slows by $50$\% (thick, transparent line, Fig.~\ref{fig:PDI_chain}d). 
The acceleration of triplet transport due to the presence of Peierls vibrations stems from faster triplet formation away from the initial triplet pair state $\ket{T_1T_2}$ (black line, Fig.~\ref{fig:PDI_chain}e), consistent with the lower population of singlet states (inset, Fig.~\ref{fig:PDI_chain}e) compared to the case without Peierls vibrations (transparent lines, Fig.~\ref{fig:PDI_chain}e).
While the current model does not account for either triplet separation or direct triplet-triplet transport, our results present an exciting view of singlet-mediated triplet transport within a fully quantum mechanical framework. We expect that future adHOPS simulations of singlet fission materials will enable deeper analysis and characterization of the singlet-mediated triplet transport mechanism.

Finally, we show that adHOPS achieves size-invariance even in calculations with simultaneous Holstein and Peierls vibrations. Mean CPU time of EP-PDI linear chain simulations stops growing by $N\approx50$, as shown in Fig. \ref{fig:PDI_chain}f.\bibnote{All CPU times were measured by running each trajectory on a single thread of an AMD EPYC 9534 64-Core Processor.} The complex structure of the open quantum system Hamiltonian, strong electronic and electronic-vibrational couplings, and existence of transiently-populated states in our model are unlike previous systems in which adHOPS has exhibited size-invariance,\cite{adHOPS,gera_simulating_2023, MesoHOPS2024, gera2025fluorescence} demonstrating that the method's improved scalability is useful in a broad range of molecular materials. Nonetheless, residual scaling beyond realistic exciton delocalization lengths indicates that it may be possible to further optimize the adaptive algorithm.

The simulations presented here elucidate key mechanistic details of SF, but a full understanding of the process requires more work. A previous Redfield calculation implicated the Peierls vibration augmenting the singlet-TT direct coupling as a primary driver of SF, but we demonstrate this is untrue in EP-PDI.\cite{Renaud2015SFPeierls} Our results are instead consistent with findings that indicate that intermolecular vibrations break anti-symmetry between effective CT-mediated couplings,\cite{Castellanos2017SFPeierls, Miyata_2017_Symmetry_Breaking,alvertis2019solventpolarity,tamura2015pistacking,Alguire2015noncondon} and reveal the importance of coherent and edge effects in such a mechanism. However, accurately representing the vibrational bath is a complex problem: intermolecular vibrations can have both Holstein and Peierls character,\cite{nagami2020vibronic} the structure of the spectral density significantly impacts dynamics,\cite{Renaud2015SFPeierls} and the assumption that each state possesses a totally uncorrelated bath is at odds with the fact that the CT and TT states physically inhabit the same molecules. Our calculations also do not explore static disorder\cite{Volek_2024_SF_TF_Rates, petelenz2016brokensymmetry}, the possibility of photoexcitation of a mixed singlet-TT state,\cite{alvertis2019solventpolarity,Miyata_2017_Symmetry_Breaking,ChanEndothermic2012} nor the role of dimensionality and entropic effects,\cite{Tempelaar2018SFPeierls,ChanEndothermic2012} all of which have been implicated in ultrafast SF. Symmetry-breaking by vibrations may also prove secondary in the landscape of material design, given previous findings that altering packing structures in pentacene is sufficient to coherently accelerate SF.\cite{zang_interference_2017} Finally, the mechanism presented here is unlikely to be universal: a previous simulation indicated that Holstein vibrations are the primary driver of SF in pentacene, even in the presence of Peierls vibrations.\cite{Tempelaar2018SFPeierls} Looking forward, exact simulations that explore SF mechanisms in more-refined models may elucidate material design principles for optimizing device performances. While the mechanism of single-exciton transport processes can be estimated from material parameters alone,\cite{Li_OSC_2021} the symmetry-breaking behavior stemming from Peierls vibrations observed here serves as a reminder that an accurate approach is necessary to understand the mechanistic underpinnings of multi-exciton processes. 

The generalized adHOPS algorithm presented in this letter is a formally exact, size-invariant ansatz that can simulate any open quantum system with a Hamiltonian expressible in the form of Eq. \eqref{eq:open_quantum_system_hamiltonian}. Our method is well-positioned to explore the complex role of Holstein and Peierls vibrations in the excited-state dynamics of molecular materials on 100s-of-nm scales. Going forward, adHOPS may be further accelerated through adjustments to the adaptive algorithm, combination with tensor contraction methods, and more-sophisticated approaches to sample long-time dynamics.

\begin{acknowledgement}
The authors thank Brian Citty, Zachary Freeman, Tarun Gera, and Alexia Hartzell for editing and code review, and Xin Chang, Sean Roberts, and Wennie Wang for helpful discussions. The table-of-contents figure was finalized by Alexia Hartzell. The authors acknowledge support from the Robert A. Welch Foundation (Grant No. N-2026-20200401), a U.S. National Science Foundation CAREER Award (Grant No. CHE-2341178), start-up funds from the University of Texas at Austin, and the Texas Advanced Computing Center (TACC) at the University of Texas at Austin (http://www.tacc.utexas.edu).
\end{acknowledgement}

\section{Data Availability Statement}
All figure, input, and analysis scripts are available on \href{https://doi.org/10.5281/zenodo.15337336}{Zenodo}.\cite{mesohops_peierls_paper_zenodo_archive} All calculations were run with an intermediate version of our code, also available in the Zenodo.

\begin{suppinfo}
Detailed description of the adaptive algorithm, derivation of adaptive error terms, detailed EP-PDI Hamiltonian and derivation of effective couplings, description of all calculation parameters, and convergence data. (PDF)

\appendix

\end{suppinfo}

\bibliography{achemso-demo}

\begin{mcitethebibliography}{48}
\bibitem[Not()]{Note-1}
All CPU times were measured by running each trajectory on a single thread of an AMD EPYC 9534 64-Core Processor.\relax
\mciteBstWouldAddEndPunctfalse
\mciteSetBstMidEndSepPunct{\mcitedefaultmidpunct}
{}{\mcitedefaultseppunct}\relax
\EndOfBibitem
\bibitem[Not()]{Note-3}
We note here that, given these vibrational relaxation effects, an equivalent material in an H-aggregate structure with a positive effective nearest-neighbor coupling $V_{NN}>0$ would have $\ket{J_1}$ as the singlet exciton ground state and the anti-symmetry $V_{HL}=-V_{LH}$ would promote, rather than impede, SF (results not shown).\relax
\mciteBstWouldAddEndPunctfalse
\mciteSetBstMidEndSepPunct{\mcitedefaultmidpunct}
{}{\mcitedefaultseppunct}\relax
\EndOfBibitem
\end{mcitethebibliography}

\end{document}